Female teachers effect on male pupils' voting behavior and preference formation.

Eiji YAMAMURA

Seinan Gakuin University, Department of Economics.

yamaei@seinan-gu.ac.jp.

Abstract

This study examines the influence of learning in a female teacher homeroom class in elementary school on pupils' voting behavior later in life, using independently collected individual-level data. Further, we evaluate its effect on preference for women's participation in the workplace in adulthood. Our study found that having a female teacher in the first year of school makes individuals more likely to vote for female candidates, and to prefer policy for female labor participation in adulthood. However, the effect is only observed among males, and not female pupils. These findings offer new evidence for the female socialization hypothesis.

1. Introduction

As political leaders, women are expected to play a critical role in reducing the gender gap in society. Female council candidates receive more preferential votes when a female mayor has recently been elected into office (Baskaran & Hessami, 2018)[1]. Exposure to female politicians during young adulthood has a long-term influence; thus increasing the probability of women working in wage employment (Priyanka, 2020)[2]. Regarding social background, the gap between male and female candidates has reduced significantly as a result of changes in social norms (King & Leigh, 2010)[3].
Moreover, changes in the political system such as women's suffrage, have had a positive impact on education and childhood (e.g., Carruthers & Wanamaker, 2015; Miller, 2008). Education contributes to the formation of subjective views and preferences (Algan, Cahuc, & Shleifer, 2013; Aspachs-Bracons, Clots-Figueras, Costa-Font, & Masella, 2008; Hryshko et al., 2011). Studies have found that these formative influences have an effect on voter turnout (Milligan, Moretti, & Oreopoulos, 2004)[4]. Moreover, being taught by female teachers also has a long-term effect later in life (Yamamura & Tsutsui, 2019)[5]. Many studies have highlighted the factors that increase the probability of females winning elections (Bauer, 2020; Hogan, 2007; King & McConnell, 2003; Lublin & Brewer, 2003; Moehling & Thomasson, 2020). However, the effect of female teachers on voting behavior has not been examined. Therefore, this study investigates how people vote for female candidates and prefer gender equalization by considering the effect of female teachers in early childhood education.

Our investigation of the effect of female teachers on pupils' voting behavior and preferences later in life is inspired by existing studies suggesting a cross-gender effect;

mothers influence their sons to prefer working women (Kawaguchi & Miyazaki, 2009) and the wives of men whose mothers worked are significantly more likely to work (Fernandez, Fogli, & Olivetti, 2004). In addition to the effect of working mothers on their sons, various types of different gender-matching effects have been observed. Having daughters transformed a man's view on women empowerment in society (e.g., Glynn & Sen, 2014; Milyo & Schosberg, 2000; Oswald & Powdthavee, 2010; Washington, 2008)[6]. These findings emphasize women's influence on men's views and preferences, which is called *female socialization.*

We independently collected individual-level data through an internet-survey directly after the election in Japan. In the survey, we asked respondents about their views on active female participation in society, and whether they voted for female candidates. We also inquired about the genders of their homeroom teachers in elementary school. Based on the data, we found the following: Generally, women were more likely than men to prefer active female participation in society and to vote for female candidates. Male pupils who had a female teacher in the first year of elementary school were more likely to vote for female candidates and prefer female social participation than those who had a male teacher. However, this effect was not observed among female pupils. This implies that female teacher–male pupil matching reduces the gender difference in voting behavior and influences one's preference to support active female involvement in society. The study aids in bridging education economics and voting behavior to provide new evidence that early childhood education facilitates a change in male pupils' views in the long-term, thus promoting *female socialization*.

The remainder of this article is organized as follows. Section 2 proposes the testable hypotheses. Section 3 describes the setting and data. Section 4 presents the empirical methodology. The estimation results and their interpretation are presented in Section 5. The final section presents some reflections and conclusions.

2. Hypothesis

Miller (2008) found that suffrage extension is positively related to public goods expenditures. Further, Miller (2008) indicated that women gaining votes led to reduced child mortality. These findings suggest that increased public goods expenditures were allocated in ways that improved child health. Carruthers and Wanamaker (2015) found that suffrage led to an increase in public school expenditures. From these findings, we can argue that women consider the well-being and future of children. During school life, children are more likely to appreciate female intention if they belong to a female teacher's homeroom class. Naturally, pupils are more likely to rely on and trust female homeroom teachers than male teachers. This gives pupils motivation to support female social participation, which persists later in life.

However, female pupils are motivated to participate in society for themselves regardless of the gender of their teachers. Hence, the effect of female teachers is observed among males, but not for female pupils. Therefore, female socialization is promoted by male pupils. Thus, we propose the following hypotheses:

*Hypothesis 1: Having a female teacher influences male pupils to vote for female candidates in the election after male students become adults.*

*Hypothesis 2: Having a female teacher influences male pupils' preference for female labor participation later in life.*

3. The setting and the data

To investigate voting behavior, we obtained individual-level data through a web-based survey in July 2016, conducted immediately after the House of Councilors election in Japan. The Nikkei Research Company was commissioned to conduct the web survey. Surveys were openly posted on the Nikkei Research, and therefore the surveys were conducted until a sufficient sample had been collected. Since we aimed to collect over 10 000 observations, the survey was active until 10 000 observations were collected. A total of 12 176 respondents were asked to complete the questionnaire.

In the questionnaire, we asked respondents whether they voted for female candidates and inquired about their views on female participation in the workplace. In addition, we obtained basic economic and demographic data such as sex, age, educational background, parental educational background, household income, job status, marital status, number of siblings, residential prefecture, and residential prefecture at six years of age. Furthermore, we gathered information about their educational experience such as if they worked and learned in groups in elementary school. This data was collected as prosocial behavior may have been facilitated by teaching practice (Algan et al., 2013). To construct the panel dataset, we conducted a follow-up survey in July 2017, which included questions about the sex of teachers in elementary school. This helped to separate the teacher's sex effect from the curriculum effect. In 2017, we gathered 9130 observations, indicating that approximately 75 % of the respondents in 2016 had also participated in 2017. We then matched the respondents of 2017 with respondents of 2016. Subsequently, a total of 7107 respondents who participated in both surveys were included in this study, where we can collected the variables of candidates sex for whom respondents voted and view about female participation. Further, some respondents could not recall the gender of their teachers in elementary school. Hence, the sample size reduced to 5024 if we were limited to observations where we gathered teachers' genders in elementary school. The total number of male and female respondents was 2595 and 2429, respectively.

Figure 1 illustrates that the sample's demographic composition is equivalent to the 2015 Japan Census composition. As for educational background in Japan, according to OECD statistics, the percentage of individuals graduating university was approximately 50.5 % in 2016[7]. In our dataset, the percentage of those who graduated from university was 56 %. Hence, to a certain extent, the dataset represents Japanese society.

Observations used for estimations are slightly reduced because some respondents did not respond to questions on variables included in the model. In Japan, there were 47 prefectures. There were 47 election districts, which were equivalent to prefectures. We asked for participants' residential prefecture to identify election districts where they voted. Out of the 47 prefectures, there were no female candidates in 15 prefectures in the 2016 election. While estimating voting behavior, we limited the sample to prefectures where female candidates stood in the 2016 election. Respondents who did not cast a vote were not included in the sample. Hence, at most, the sample used for the voting behavior and female participation was 2192 and 3350, respectively.

Table 1 provides definitions of the key variables and their descriptive statistics, based on the sample used for the estimation of the view of female participation. In the first year, 81 % of the pupils were assigned to a female teacher class. This rate monotonically declined to 40 % in the sixth year. As is well known, in Japan, women teachers tend to teacher lower grades compared to male teachers. This is because the workload is larger in higher grades. For instance, teachers are obliged to lead higher grade students to overnight school excursions. Commonly, women teachers balance housework along with their work as teachers. Therefore, they avoid teaching a higher-grade class.

Parents cannot choose the gender of the teacher, especially in the first grade. This means that the random assignment criteria of natural experiments has been met (Yamamura & Tsutsui, 2019). Teachers are acquainted with pupils' characteristics and dispositions while teaching them and observing their behavior in school. In higher grades, schools have more information regarding matching between pupils and teachers. If a conflict arises, the pupil is assigned to a presumably more suitable teacher in the next year. If pupils were inappropriately matched with a female teacher class in the past, they might be assigned to a male teacher class. In other words, the assignment to a female teacher class in higher grades seems to be determined by accumulated information about the compatibility between teacher–pupil genders. Therefore, the assignment to a women's class in first grade is more random and exogenous than other grades. Hence, the first-grade assignment is free from selection bias.

As explained, assignment of a class was randomized in the first year. However, the probability of being assigned to a female teacher class may vary according to the female teacher ratio in the area where respondents resided in the year of entering school. From official surveys, we gathered the number of both male and female teachers for 47 prefectures in different years, which enabled us to calculate the female teacher rate[8]. We also gathered information about respondents' residential prefectures at six years of age. We then matched the ratio of female teachers with the respondents by considering their years of entering school and their respective prefectures at the age of six[9]. Table 1 shows that the female teacher ratio ranged between 0.24 and 0.73, indicating a wide variation of the probability of being assigned to female teacher class according to time and place. From our original data, we calculated years of being assigned to a female teacher class during the elementary school period, which ranged between 0 and 6 as there are six grades in the elementary school of Japan. To compare the relationship between the female teacher ratio and the probability of being assigned to a female teacher class, we calculated these standardized values which are illustrated in Figure 2. Figure 2 shows that both years of female teacher class and female teacher ratio increased when respondents were younger. Figure 3 shows a comparison of years of being in a female teacher class between males and females. We observed a similar trend of being assigned to a female teacher class between them.

In Table 2, we compared the ratio of belonging to the female teacher class between male and female pupils. With the exception of the second grade, there was no statistically significant difference. In the second grade, the teacher gathered information about the characteristics of the pupils through teaching them in the first grade. Therefore, some pupils are selectively assigned to a male teacher class in the second grade if they are more likely to be better suited to male teachers than female teachers. However, in general, there is no bias when pupils are assigned to a female teacher class. Moreover, we also checked the female teacher ratio of residential prefectures and the probability of being assigned to a female teacher class. Table 3 shows the mean difference test of the female teacher ratio in residential prefectures between the group in a female teacher class and the group in a male teacher class in each grade. For male respondents, the female teacher ratio was higher by approximately 0.03 points for those assigned to a female teacher class than for those in a male teacher class regardless of grades. The statistical significance level was 1 % in all grades. A similar tendency was also observed for female respondents. These observations suggest that the female teacher rate in the residential prefecture increased the probability of being assigned to a female teacher class despite random assignment.

It is plausible that younger respondents are more able to recall the teacher's sex in elementary school, causing bias. However, as illustrated in Figure 4, response rates for questions about teachers' sex in elementary school are almost 80 % and are almost the same in each cohort group. Therefore, bias is unlikely to have occurred. According to a 2015 survey on information technology, over 90 % of the working-age population in Japan are web users. Therefore, selection bias for web users does not need to be considered.[10]

According to the definition of key variables in Table 1, *Vote woman* is a dummy variable that accepts 0 or 1 while *Support woman* ranges between 1 and 5. The larger these variables, the more respondents are likely to support active female participation in society. To compare these variables, we standardized them in Figure 5. A cursory examination of Figure 5 shows that female respondents are more likely to support women's roles that are in line with intuition. The difference in *Support woman* between male and female respondents was larger than that in *Vote woman.* We interpreted this as follows: female candidates have political opinions, with varying beliefs on females' roles in society. Hence, some female candidates may have a traditional view and are therefore less likely to support women's active role in society as compared to male candidates in the same election district. Therefore, *Support woman* more directly captures the respondents' views than *Vote woman*.

Figures 6 and 7 illustrate the differences in *Vote* and *Support woman* between those assigned to female and male teacher classes in each grade, respectively. Figure 6 shows that overall, male pupils assigned to a female teacher's class in lower grades were more likely to vote for female candidates. In contrast, female pupils assigned to a male teacher class were more likely to vote for female candidates. From Figure 7, we see that both male and female pupils assigned to a female class were more likely to support women's participation in the workplace. The difference in these variables between female and male teachers' classes is the largest in first grade and declines as pupils are promoted. These findings suggest that female teachers may facilitate a positive view on female roles in society among pupils when they become adults. This effect is significant in the first grade.

4. Empirical methodology

Our baseline model assesses the influence of a female teacher homeroom class in elementary school on pupils' voting behavior and views on women's role later in life. The estimated function takes the following form:

*Vote woman* $_i$ (or *support woman* $_i$)= $\alpha_0 + \alpha_1$ *Female teacher in first year* +

$\alpha_2$ *Years of female teacher from the second to-sixth year*$_i$ + X$'_i$ B + u $_i$.

*Vote woman* $_i$ or *Support woman* $_i$ is the dependent variable. *Vote woman* is a dummy variable that accepts 0 or 1, and thus the Probit model was used. *The Support woman* variable ranges between 1 and 5, and thus the OLS model was used. The key independent variable is *Female teacher in first year* because it captures the random assignment to the female teacher class. Its coefficient has a positive sign if female teachers in the first year influence pupils to vote for female candidates and support women's active participation in society later in life. In addition to the full-sample estimation, we use sub-samples divided by the respondents' gender to examine the teacher–pupil gender-matching effects. *Female teacher in the first year* shows a significant positive sign only for the male sample, if a different gender-matching effect exists (e.g., Oswald & Powdthavee, 2010; Washington, 2008). To control for the influence of female teacher class in higher grades, we included *Years of female teachers from the second to sixth year*, which aggregated years in higher grades during the elementary school period. In alternative specification, instead of *Years of female teacher from the second to sixth year*, we simply added five dummies of the female teacher class.

In addition, the vector of the control variables is denoted by $X_i$ and the vector of the estimated coefficients is denoted by B. As control variables, we added the female teacher ratio of the respondent's residential prefecture at the respondent's school age. Further, we added the number of female candidates in the respondent's election district. In addition, the control variables were seven dummies for educational background as a proxy for the quantity of education, age, 17 income dummies, and 19 occupation dummies. We also controlled for variables such as group work and pro-competition curricula because specific educational features such as teaching practices could have influenced pupils' preferences and world views (e.g., Algan et al., 2013; Aspachs-Bracons et al., 2008; Milligan et al., 2004). It is plausible that family conditions also influenced the formation of preferences. Parents' education levels are controlled for by including the father's and mother's educational attainment dummies. Further, family composition is an important factor that affects views on social and economic issues (e.g., Borrel-Porta, Costa-Font, & Philipp, 2019; Oswald & Powdthavee, 2010; Washington, 2008). Therefore, the number of siblings and the dummies of marital status are included separately. The estimation results for these control variables were not reported. However, these variables are included in all estimations.[11]

5. Estimation results

Table 1 shows that the sample included the election district without female candidates. Hence, to examine voting behavior, we used a sub-sample that excluded observations without female candidates. Hence, the sample size for the estimation of *Support woman* is larger than that of *Vote woman*.

We began by examining the results of *Vote woman* estimations. Table 4 shows that the coefficient of *"Female teacher in first year"* was positive in all columns. We observed statistical significance for the male sample, but not for the female sample. Years of female teacher in higher grades and other female teacher dummies did not show a significant positive sign, with the exception of *Female teacher in sixth year* in column (4). These results are consistent with our prediction. Its marginal effect is 0.10–0.11 for the male sample which indicates that a man was 10 or 11 % more likely to vote for female candidates if he was assigned to a female teacher class in the first year of elementary school. These results support *Hypothesis 1*. Besides this, other variables did not exhibit statistical significance.

Regarding the results of *Support woman,* Table 5 shows that the coefficient of *Female teacher in first year* was positive in all columns. We observed statistical significance for the male sample as well as the sample composed of male and female respondents, but not for the female sample. The value of the coefficient was 0.15 in column (4), indicating that a male's support of women's participation in the workplace was greater by 0.15 points on the 5 point scale if he was assigned to a female teacher class in the first year of elementary school. We observed significant negative correlations for *Years of female teacher from the second to sixth year* and *Female teacher in fifth year* for males. This is not consistent with our hypotheses. However, these results may have resulted from bias because information about previous years in elementary school affects class assignment. These results are consistent with *Hypothesis 2.* As shown by Figure 5, females were positively statistically significant at the 1 % level. Thus, they are more inclined to support their active role in society.

From our analysis, we conclude that female teachers influenced the world view of male pupils directly after entering elementary school; thus, driving them to support female participation in politics as well as in the labor market.

6. Conclusion

This study explored how education reduces the gender gap in society. In particular, we focus on the effect of female teachers on the formation of male pupils' views in early childhood education. For this purpose, we employed a quasi-natural experiment of teacher–student random gender matching in first-grade elementary schools in Japan. Using independently collected individual-level data directly before the House of Councilors election, we found that males are more likely to vote for female candidates and to prefer policy for female labor participation if they belonged to a female teacher's homeroom class in the first grade of elementary school. However, this effect was not observed for female respondents.

From these findings, we argue that exposure to the opposite gender in early childhood leads males to have female role models. This holds true not only within one's family (Fernandez et al., 2004; Kawaguchi & Miyazaki, 2009), but also in school. Therefore, female teachers play roles similar to that of a mother for male pupils in early childhood. This study contributes to the endeavor of bridging education and voting behaviors to support the female socialization hypothesis.

Like all empirical work, there are some limitations of this study. The key independent variable (female teacher dummy) is a recall variable that possibly suffers from measurement error bias[12]. Further, the dependent variable of voting for females is binary, which may have resulted in a significant loss of information that could have been captured with a continuous dependent variable. However, these limitations do not undermine the study's value. In contrast, they highlight that further work on the effects of teachers' genders on voting behavior is warranted.

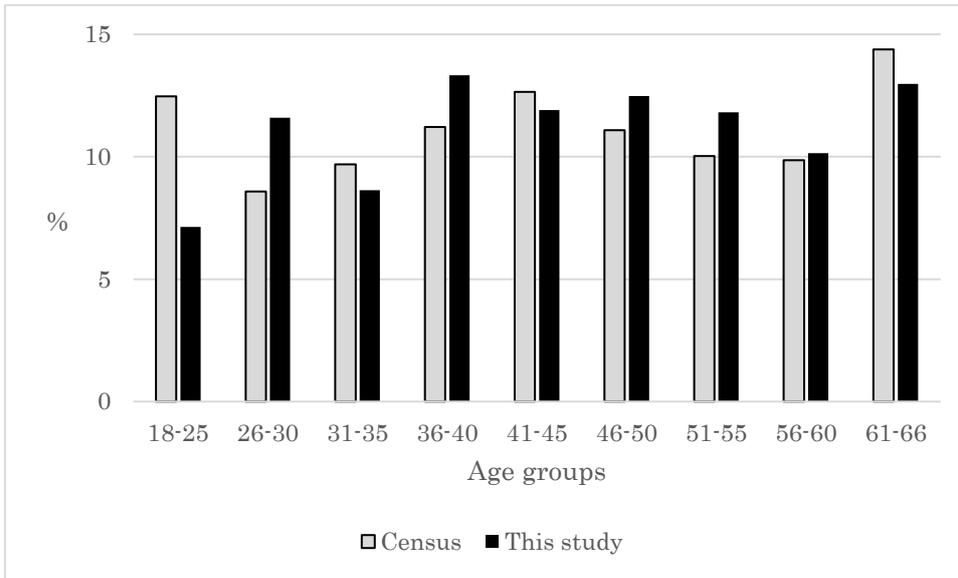

Figure 1. Comparison between of the age distribution of our sample and the 2015 Japan population census.

Source: Statistics Bureau, Ministry of Internal Affairs and Communications (2015). 2015 Japan Population Census. https://www.e-stat.go.jp/stat-search/files?page=1&layout=datalist&toukei=00200521&tstat=000001080615&cycle=0&tclass1=000001089055&tclass2=000001089056&result_page=1&second=1&second2=1 (accessed on February 18, 2018).

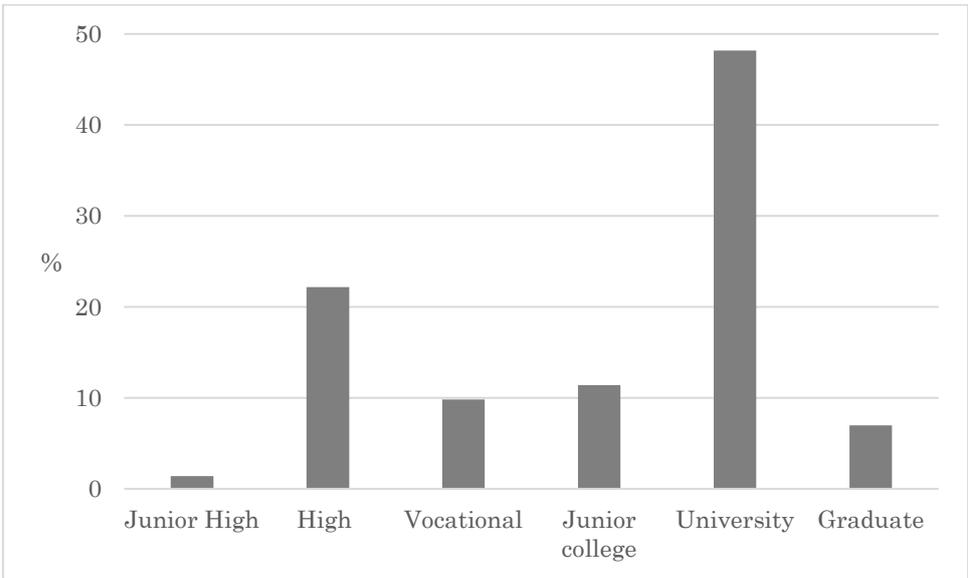

Figure 2. Comparison of educational background.

Note: "High" means high school. "Vocational" means vocational school which have been entered after graduating from high school. "Graduate" means graduate school.

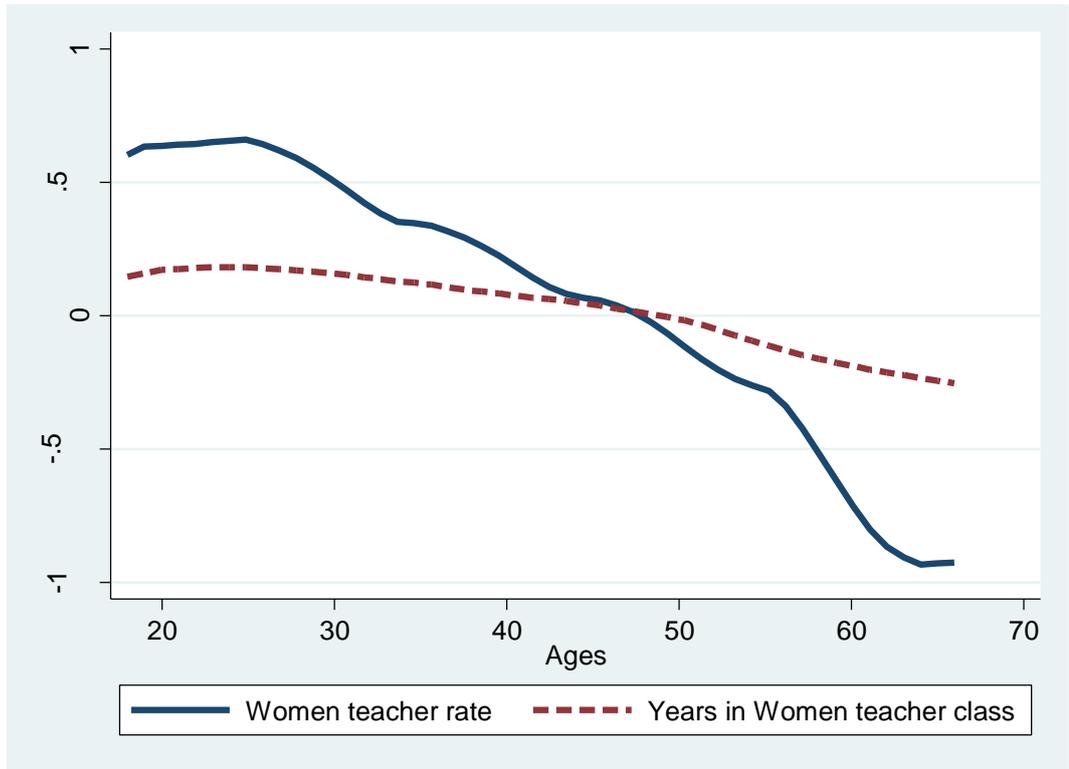

Figure 2.
"Years of female teacher class" versus "Female teacher rate in residential prefecture during elementary school."

Source: Female teacher rate collected from "Report on School Basic Survey (various years)."

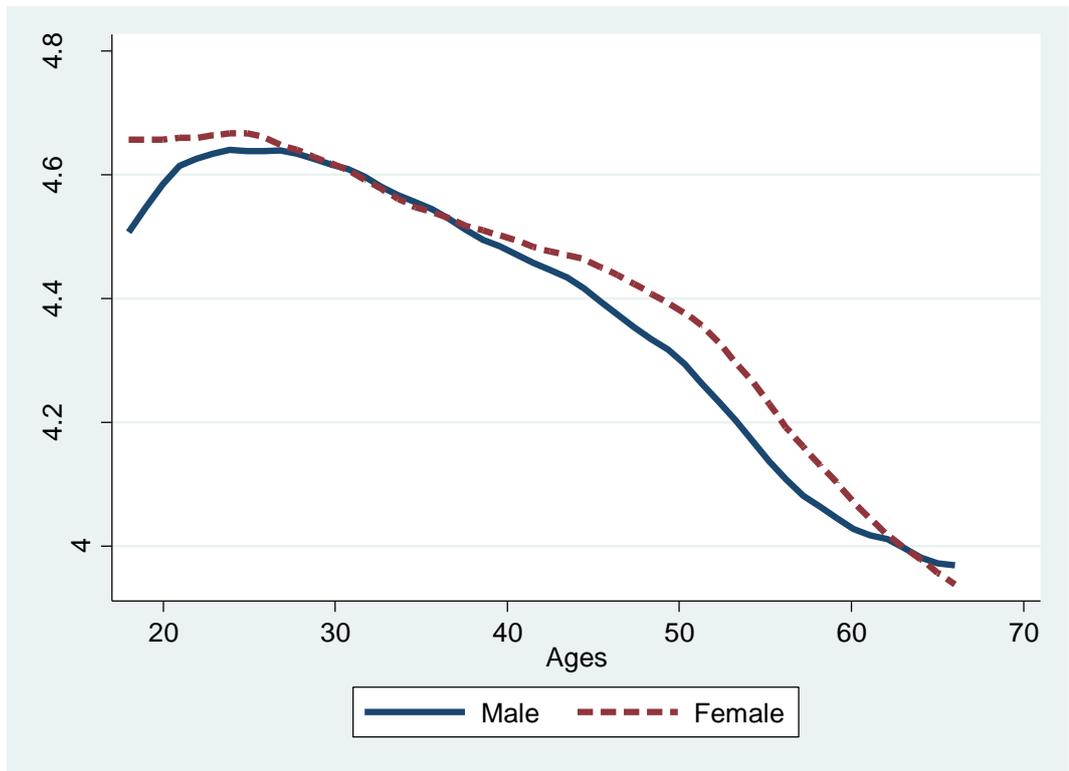

Figure 3. Comparison between respondent's genders: Years in female teacher class during elementary school

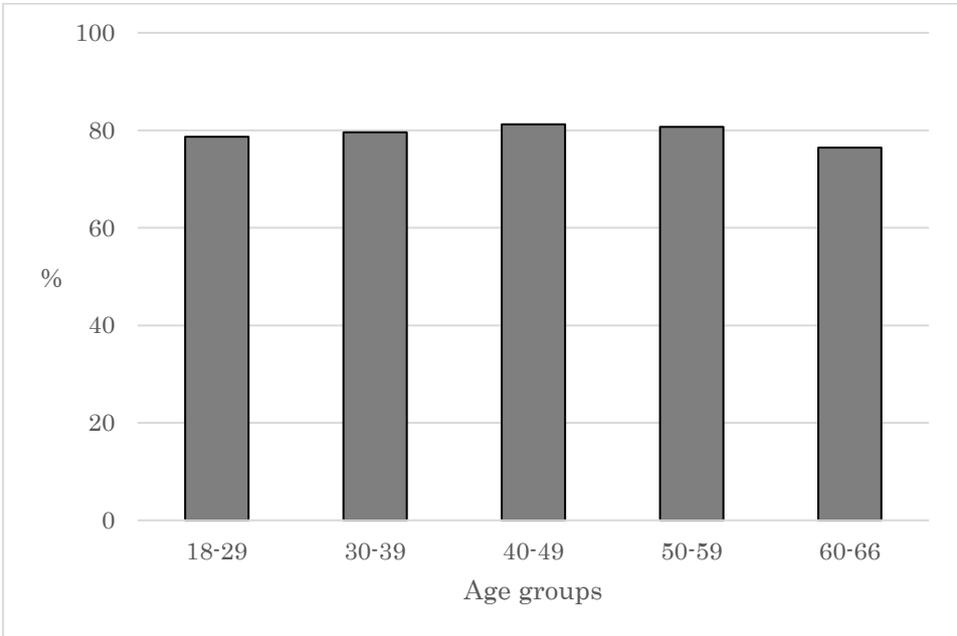

Figure 4. Rate of replying to questions about sex of homeroom teacher in elementary school in each age group.

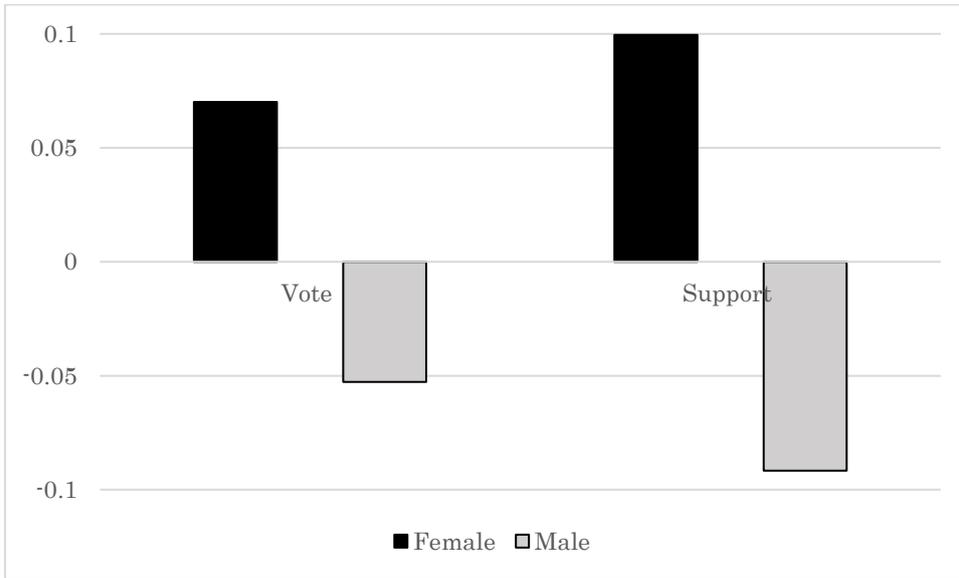

Figure 5. Comparison of standardized variables between male and female respondents

Note: Vote is the mean of standardized values of *Vote woman*
Support is the mean of standardized values of *Support woman*

Male sample

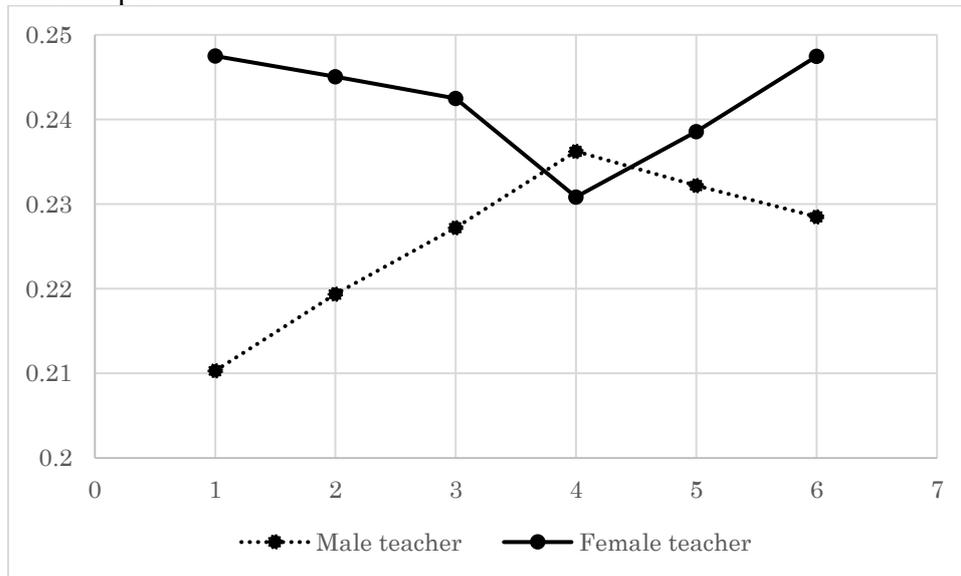

Female sample

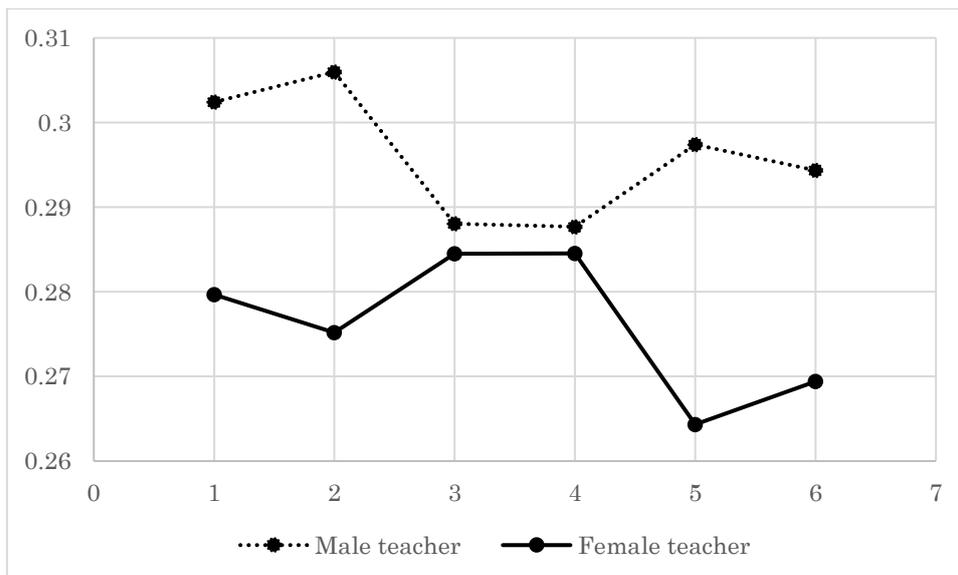

Figure 6. *Vote woman* in adulthood between people who belonged to female and male classes in each grade in elementary school

Male sample

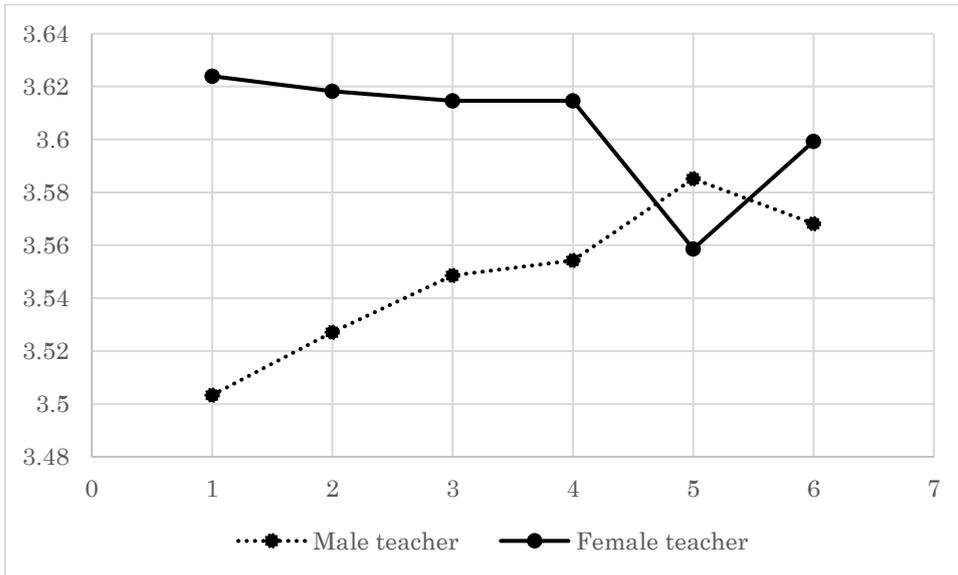

Female sample

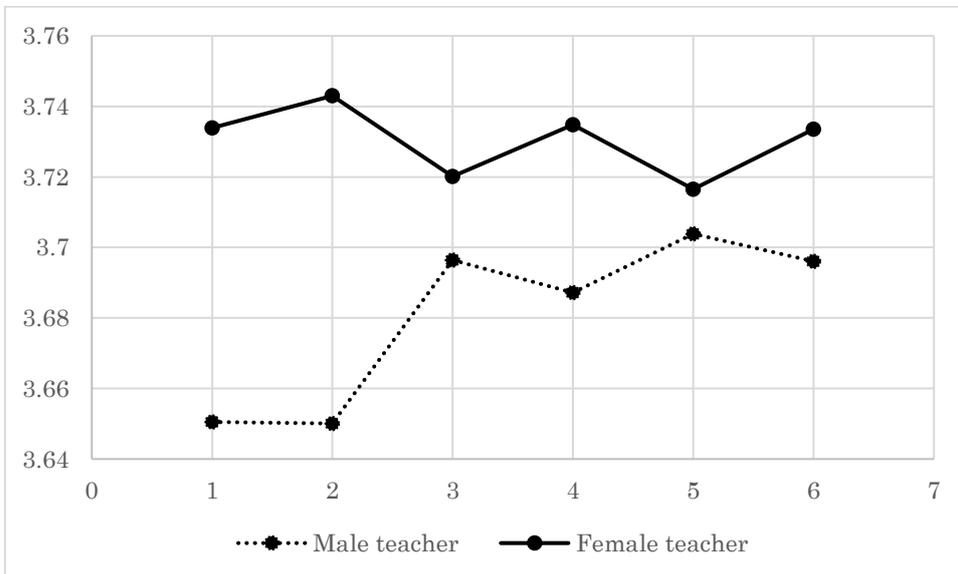

Figure 7. *Support woman* in adulthood between people who belonged to women's and men's classes in each grade in elementary school

Table 1. Definitions of key variables and their basic statistics

| Variables | Definition | Mean | Standard deviation | Min. | Max. |
|---|---|---|---|---|---|
| *Dependent Variables* | | | | | |
| *Vote woman* | Equals 1 if the respondent voted for a female candidate, 0 otherwise | 0.28 | 0.45 | 0 | 1 |
| *Support woman* | Do you agree that government should create an economic and social environment in which women are able to fully exhibit their ability and actively participate in workplaces? 1 (strongly disagree)–5 (strongly agree) | 3.70 | 0.94 | 1 | 5 |
| *Female teacher in first year.* | Equals 1 if class teacher is female at the first grade in elementary school, 0 otherwise | 0.81 | 0.39 | 0 | 1 |
| *Female teacher in second year.* | Equals 1 if class teacher is female at the second grade in elementary school, 0 otherwise | 0.73 | 0.44 | 0 | 1 |
| *Female teacher in third year.* | Equals 1 if class teacher is female at the third grade in elementary school, 0 otherwise | 0.58 | 0.49 | 0 | 1 |
| *Female teacher in fourth year.* | Equals 1 if class teacher is female at the fourth grade in elementary school, 0 otherwise | 0.51 | 0.39 | 0 | 1 |
| *Female teacher in fifth year.* | Equals 1 if class teacher is female at the fifth grade in elementary school, 0 otherwise | 0.40 | 0.49 | 0 | 1 |
| *Female teacher in sixth year.* | Equals 1 if class teacher is female at the sixth grade in elementary school, 0 otherwise | 0.39 | 0.49 | 0 | 1 |
| *Years of female teacher from second to sixth year* | Total years of female teacher class between second and sixth grades. | 2.61 | 1.39 | 0 | 5 |
| *Female teacher rate* | The ratio of female teachers in respondents' residential areas when they were in the elementary school. | 0.57 | 0.10 | 0.24 | 0.73 |
| *Number of female candidates* | Number of female candidate in respondents' election district in the 2016 election. | 3.05 | 2.04 | 1 | 7 |
| *Schooling years* | Respondent's schooling years | 14.8 | 1.91 | 6 | 18 |
| *Age* | Respondents' age | 44.0 | 12.5 | 18 | 67 |
| *Women* | Equals 1 if the respondent is a woman, 0 otherwise | 0.50 | 0.50 | 0 | 1 |

Note: Apart from the job dummies indicated, 13 other job dummies were included in the estimation model: (1) Chief executive officer, (2) Temporary employee, (3) Public officer, (4) Specialists (lawyers, accountants), (5) Self-employment, (6) SOHO (Small Office Home Office), (7) Part-time worker, (8) Outside worker, (9) House worker, (10) University student, (11) High school student, (12) Unemployed or retired, (13) Other worker.

Table 2. Mean difference test of female teacher dummy in each grade.

|  | (1) Female respondents | (2) Male respondents | (1)-(2) |
|---|---|---|---|
| *First year.* | 0.82 | 0.80 | 0.02 |
| *Second year.* | 0.75 | 0.72 | 0.03** |
| *Third year.* | 0.59 | 0.58 | 0.01 |
| *Fourth year.* | 0.52 | 0.40 | 0.02 |
| *Fifth year.* | 0.40 | 0.40 | 0.004 |
| *Sixth year.* | 0.39 | 0.39 | 0.003 |

Notes: ** denotes statistical significance at the 5 % level

Table 3. Mean difference test of the female teacher ratio in residential prefecture between female teacher class and male teacher class groups in each grade.

|  | Female respondents | | | | Male respondents | | |
| --- | --- | --- | --- | --- | --- | --- | --- |
|  | (1) Female teacher | (2) Male teacher | (1)-(2) | | (1) Female teacher | (2) Male teacher | (1)-(2) |
| *First year.* | 0.57 | 0.54 | 0.03*** | | 0.57 | 0.53 | 0.03*** |
| *Second year.* | 0.57 | 0.55 | 0.03*** | | 0.57 | 0.54 | 0.02*** |
| *Third year.* | 0.57 | 0.55 | 0.03*** | | 0.57 | 0.54 | 0.02*** |
| *Fourth year.* | 0.58 | 0.55 | 0.02*** | | 0.57 | 0.54 | 0.03*** |
| *Fifth year.* | 0.58 | 0.55 | 0.02*** | | 0.58 | 0.55 | 0.03*** |
| *Sixth year.* | 0.58 | 0.55 | 0.03*** | | 0.57 | 0.54 | 0.03*** |

Notes: *** denotes statistical significance at the 1 % level

Table 4. Estimation results: Dependent variable is *Vote woman* (Probit estimation). Sample limited to residential areas with female candidates.

| | All | | Male | | Female | |
|---|---|---|---|---|---|---|
| | (1) | (2) | (3) | (4) | (5) | (6) |
| Female teacher in first year. | 0.06** (0.02) | 0.05* (0.03) | 0.11*** (0.03) | 0.10*** (0.03) | 0.01 (0.04) | 0.006 (0.04) |
| Years of female teachers from second to sixth year | −0.0002 (0.01) | | 0.01 (0.01) | | −0.01 (0.01) | |
| Female teacher in second year. | | 0.02 (0.02) | | 0.03 (0.04) | | −0.001 (0.04) |
| Female teacher in third year. | | −0.004 (0.02) | | −0.01 (0.03) | | −0.01 (0.03) |
| Female teacher in fourth year. | | 0.002 (0.02) | | 0.01 (0.03) | | 0.01 (0.03) |
| Female teacher in fifth year. | | −0.03 (0.03) | | −0.05 (0.03) | | −0.01 (0.04) |
| Female teacher in sixth year. | | 0.03 (0.03) | | 0.07* (0.03) | | −0.03 (0.05) |
| Female teacher rate | 0.06 (0.19) | 0.05 (0.20) | 0.31 (0.28) | 0.31 (0.29) | −0.19 (0.17) | −0.18 (0.18) |
| Number of female candidates | 0.01 (0.01) | 0.01 (0.01) | −0.001 (0.01) | −0.001 (0.01) | 0.01 (0.01) | 0.01 (0.01) |
| Female | 0.01 (0.03) | 0.01 (0.03) | | | | |
| Pseudo R-squared | 0.02 | 0.02 | 0.04 | 0.04 | 0.04 | 0.04 |
| Observations | 2192 | 2192 | 1200 | 1200 | 980 | 980 |

Notes: *** and ** denote statistical significance at the 1 % and 5 % levels, respectively. Values without parentheses are the marginal effects. Values in parentheses are standard errors, clustered by prefectures. The sample is restricted to areas with female candidates. In all columns, various control variables such as dummies for education method (Assign a value of 1 if there was a task in which students worked together as a group at elementary school; if not, assign a value of 0. Assign a value of 1 if there were running races during sporting events at elementary school and teachers ranked

the finishing order; if not, assign a value of 0), schooling years, ages, number of children, household income, marital status dummies, job dummies, father's and mother's educational attainment dummies, and a constant are included. However, these estimates have not been reported.

Table 5. Estimation results of the baseline model: Dependent variable is *Support woman* (OLS estimation).

|  | All | | Male | | Female | |
|---|---|---|---|---|---|---|
|  | (1) | (2) | (3) | (4) | (5) | (6) |
| *Female teacher in first year.* | 0.09** (0.04) | 0.10** (0.04) | 0.10* (0.06) | 0.15** (0.06) | 0.07 (0.05) | 0.04 (0.06) |
| *Years of female teacher from second to sixth year.* | −0.01 (0.01) |  | −0.04** (0.02) |  | 0.01 (0.01) |  |
| *Female teacher in second year.* |  | −0.02 (0.04) |  | −0.12 (0.08) |  | 0.09* (0.05) |
| *Female teacher in third year.* |  | −0.02 (0.03) |  | −0.04 (0.04) |  | −0.01 (0.05) |
| *Female teacher in fourth year.* |  | 0.006 (0.03) |  | −0.03 (0.05) |  | 0.04 (0.05) |
| *Female teacher in fifth year.* |  | −0.07* (0.04) |  | −0.11** (0.05) |  | −0.02 (0.07) |
| *Female teacher in sixth year.* |  | 0.06 (0.04) |  | 0.05 (0.05) |  | 0.06 (0.06) |
| Female teacher rate. | −0.27* (0.15) | −0.25 (0.15) | −0.23 (0.26) | 0.25 (0.25) | −0.23 (0.17) | −0.23 (0.18) |
| Female | 0.11*** (0.03) | 0.10*** (0.03) |  |  |  |  |
| R-squared | 0.05 | 0.05 | 0.06 | 0.07 | 0.06 | 0.06 |
| Observations | 3350 | 3350 | 1709 | 1709 | 1641 | 1641 |

Notes: ***, **, and * denote statistical significance at the 1 %, 5 %, and 10 % levels, respectively. Numbers in parentheses are standard errors clustered by prefectures. In all columns, the control variables in Table 4 are included. However, these estimates have not been reported

[1] Women incumbents did not increase the number of new women candidates in their districts, and did not increase the vote share new women candidates received (Clayton & Tang, 2018). Election of an additional female candidate results in fewer newly participating female candidates in the following elections (Kuliomina, 2018).

[2] Female mayors are less likely to engage in corruption compared to male mayors (Brollo & Troiano, 2016).

[3] Parties tend to nominate female candidates to poorer positions on the ballot (Esteve-Volart & Bagues, 2012). Meanwhile, there is no systematic bias against female candidates (Frederick & Streb, 2008). Introduction of a 50 % gender quota in candidate lists increased the probability that voters will choose women candidates (Bonomi, Brosio, & Di Tommaso, 2013).

[4] In college, compared with male advisors, female advisors are less likely to recommend mathematics as a major to students (Thompson, 2017).

[5] Early childhood education is observed to change students' life paths (e.g., Heckman et al., 2010a, 2010b, 2013).

[6] Cronqvist and Yu (2017) found that corporate social responsibility increases after a firm's chief executive officer has a daughter.

[7] Source was from the official web-site of OECD:
https://data.oecd.org/eduatt/adult-education-level.htm (accessed on April 19, 2019).
In other words, data of OECD show the percentage of population with tertiary education.
Population with tertiary education is defined as those having completed the highest level of education, by age group. Therefore, we consider tertiary education as an undergraduate degree.

[8] "Report on School Basic Survey (various years)"

[9] Those who entered elementary school in 1963, in the Kyoto prefecture. We matched the ratio of female teachers in Kyoto, and 1965 is the year closest to 1963 in our sample.

[10] Data is available from the official website of the Statistics Bureau, Ministry of Internal Affairs and Communications http://www.soumu.go.jp/johotsusintokei/statistics/statistics05.html (accessed on April 5, 2018).

[11] The results for the control variables are available from the corresponding author upon request.

[12] However, since measurement errors imply a downward bias on our estimates, our findings would likely have been even more robust had we been able to take the measurement error bias into account.